 \documentclass[final,5p,times,twocolumn]{elsarticle}

\usepackage{amssymb}
\journal{Nucl. Instrum. Methods. Phys. Res. A}

\begin{document}

\begin{frontmatter}

%% Title, authors and addresses

% use the tnoteref command within \title for footnotes;
%% use the tnotetext command for theassociated footnote;
%% use the fnref command within \author or \address for footnotes;
%% use the fntext command for theassociated footnote;
%% use the corref command within \author for corresponding author footnotes;
%% use the cortext command for theassociated footnote;
%% use the ead command for the email address,
%% and the form \ead[url] for the home page:
%% \title{Title\tnoteref{label1}}
%% \tnotetext[label1]{}
%% \author{Name\corref{cor1}\fnref{label2}}
%% \ead{email address}
%% \ead[url]{home page}
%% \fntext[label2]{}
%% \cortext[cor1]{}
%% \affiliation{organization={},
%%             addressline={},
%%             city={},
%%             postcode={},
%%             state={},
%%             country={}}
%% \fntext[label3]{}

  \title{Measurement of emission spectrum for gaseous argon electroluminescence in visible light region from 300 to 600 nm}

%% use optional labels to link authors explicitly to addresses:
%% \author[label1,label2]{}
%% \affiliation[label1]{organization={},
%%             addressline={},
%%             city={},
%%             postcode={},
%%             state={},
%%             country={}}
%%
%% \affiliation[label2]{organization={},
%%             addressline={},
%%             city={},
%%             postcode={},
%%             state={},
%%             country={}}

\author[label1]{Kazutaka Aoyama}
\author[label1,label2]{Masato Kimura}
\author[label1]{Hiroyuki Morohoshi}
\author[label1]{Tomomasa Takeda}
\author[label1]{Masashi Tanaka\corref{mt}}
\author[label1]{Kohei Yorita}
\cortext[mt]{masashi.tanaka@aoni.waseda.jp}

\affiliation[label1]{organization={Waseda University},%Department and Organization
            addressline={3-4-1 Okubo}, 
            city={Shinjyuku},
            postcode={169-8555}, 
            state={Tokyo},
            country={Japan}}

\affiliation[label2]{organization={AstroCeNT, Nicolaus Copernicus Astronomical Center of the Polish Academy of Sciences},%Department and Organization
            addressline={ul. Rektorska 4}, 
            city={Warsaw},
            postcode={00-614}, 
            country={Poland}}

\begin{abstract}
  %% Text of abstract
  A double-phase Ar detector can efficiently identify particles and reconstruct their positions. However, the properties of electroluminescence (EL) for secondary light emission in the gas phase are not fully understood. Earlier studies have explained the EL process using an ordinary EL mechanism because of an Ar excimer; however, this mechanism does not predict the emission of visible light (VL). Recent measurements have demonstrated VL components in Ar gas EL, to explain which a new mechanism called neutral bremsstrahlung (NBrS) was proposed. In this study, we investigated gaseous Ar EL in the VL region from 300 to 600 nm at room temperature and normal pressure using a gaseous time projection chamber (TPC). The secondary emission light from the TPC luminescence region was dispersed using a spectrometer. The observed spectrum was interpreted using the ordinary EL and NBrS models, and the effect of nitrogen impurities is discussed herein.

\end{abstract}

%%Graphical abstract
%%\begin{graphicalabstract}
%\includegraphics{grabs}
%%\end{graphicalabstract}

%%Research highlights
%%\begin{highlights}
%%\item Research highlight 1
%%\item Research highlight 2
%%\end{highlights}

\begin{keyword}
%% keywords here, in the form: keyword \sep keyword
Liquid argon detectors \sep Dark matter detectors \sep Neutral bremsstrahlung \sep Proportional electroluminescence
%% PACS codes here, in the form: \PACS code \sep code

%% MSC codes here, in the form: \MSC code \sep code
%% or \MSC[2008] code \sep code (2000 is the default)

\end{keyword}

\end{frontmatter}

%%\linenumbers

%% main text

\section{Introduction}
Liquid Ar is widely used as a radiation detector in particle physics experiments \cite{ATLAS:1996ab,Agnes:2014bvk,Abi:2020wmh}. The most important feature of a liquid Ar detector is that it induces two signals: scintillation photons and ionization electrons. The combination of these signals provides useful information for applications such as particle identification and position reconstruction. In particular, the double-phase argon detector technique is employed in experiments for dark matter detection \cite{Agnes:2014bvk}. The ionization electrons drifting upwards are extracted from the liquid to gas phases under a high electric field (few kV/cm). Subsequently, the ionization electrons emit proportional electroluminescence (EL) by scattering on the gaseous Ar (GAr) atoms.

\begin{figure}[h]
\centering\includegraphics[width=0.9\columnwidth]{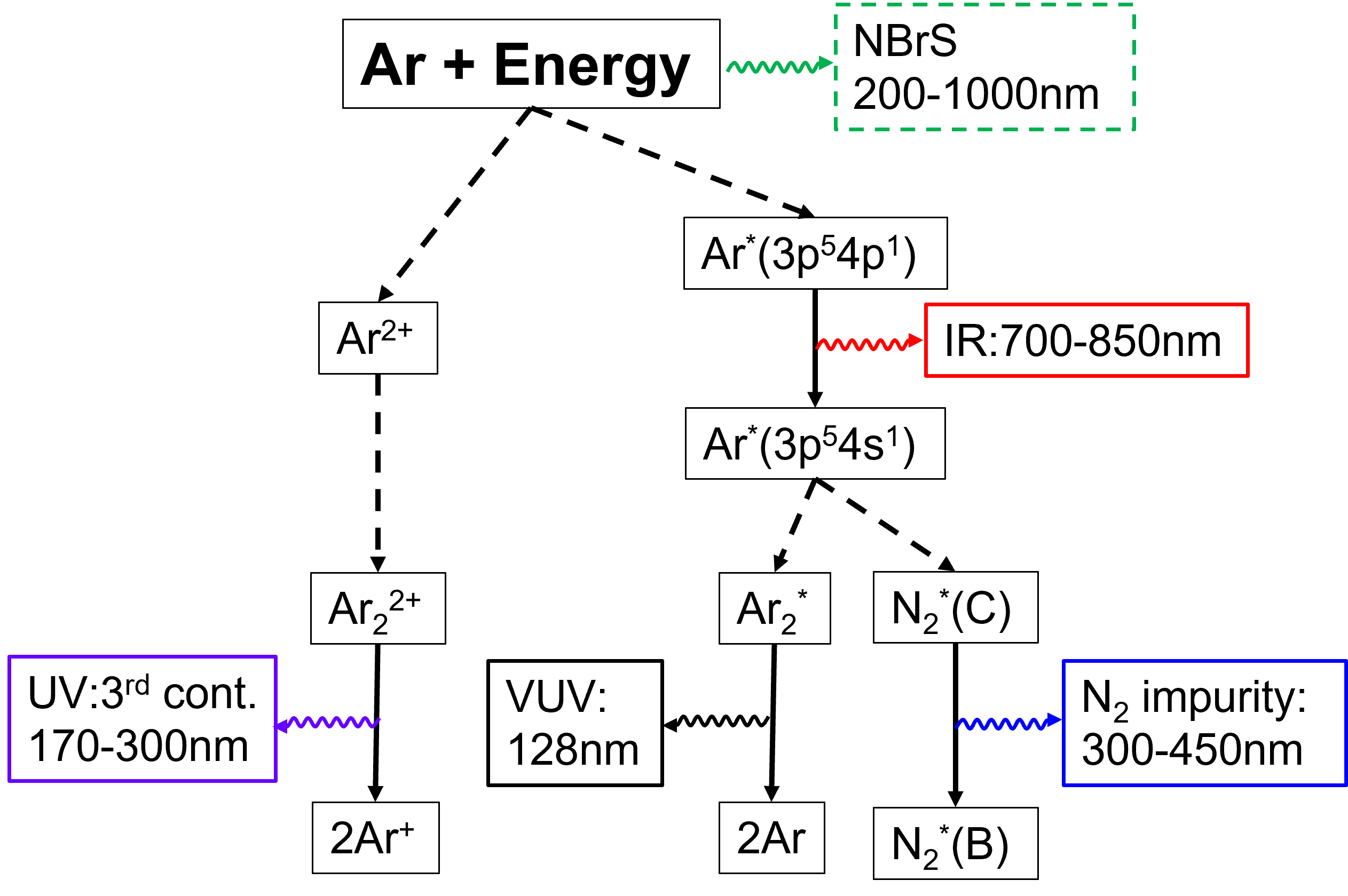}
\caption{Simplified diagram of the electroluminescence (EL) mechanism of the Ar-N mixture gas.}
\label{Fig:mech}
\end{figure}

Figure~\ref{Fig:mech} is a simplified diagram of the EL mechanism of GAr.
It describes three emission processes for pure GAr (hereinafter called ``ordinary EL'').
Two of them are well-known processes: 128-nm vacuum ultraviolet (VUV) light because of transitions of Ar excimers,
\begin{equation}
  \mathrm{Ar_2^*}  \rightarrow \mathrm{2Ar + h\nu~(VUV: 128 nm)},
\end{equation}
and 700--850 nm infrared light (IR) owing to Ar atomic emission,
\begin{equation}
  \mathrm{Ar^*(3p^54p^1)}  \rightarrow \mathrm{Ar^*(3p^54s^1) + h\nu~(IR: 700-850 nm)}.
\end{equation}
The third one is a 170--300 nm ultraviolet light (UV) emission that occurs due to the transitions of Ar molecular ions (hereinafter called ``third continuum'') \cite{Klein:1981},
\begin{equation}
  \mathrm{Ar_2^{2+}} \rightarrow \mathrm{2Ar^+ + h\nu~(UV: 170-300 nm)},
  \label{Eq:uv}
\end{equation}
which is not well understood.

\begin{figure}[!h]
  \centering\includegraphics[width=0.6\columnwidth]{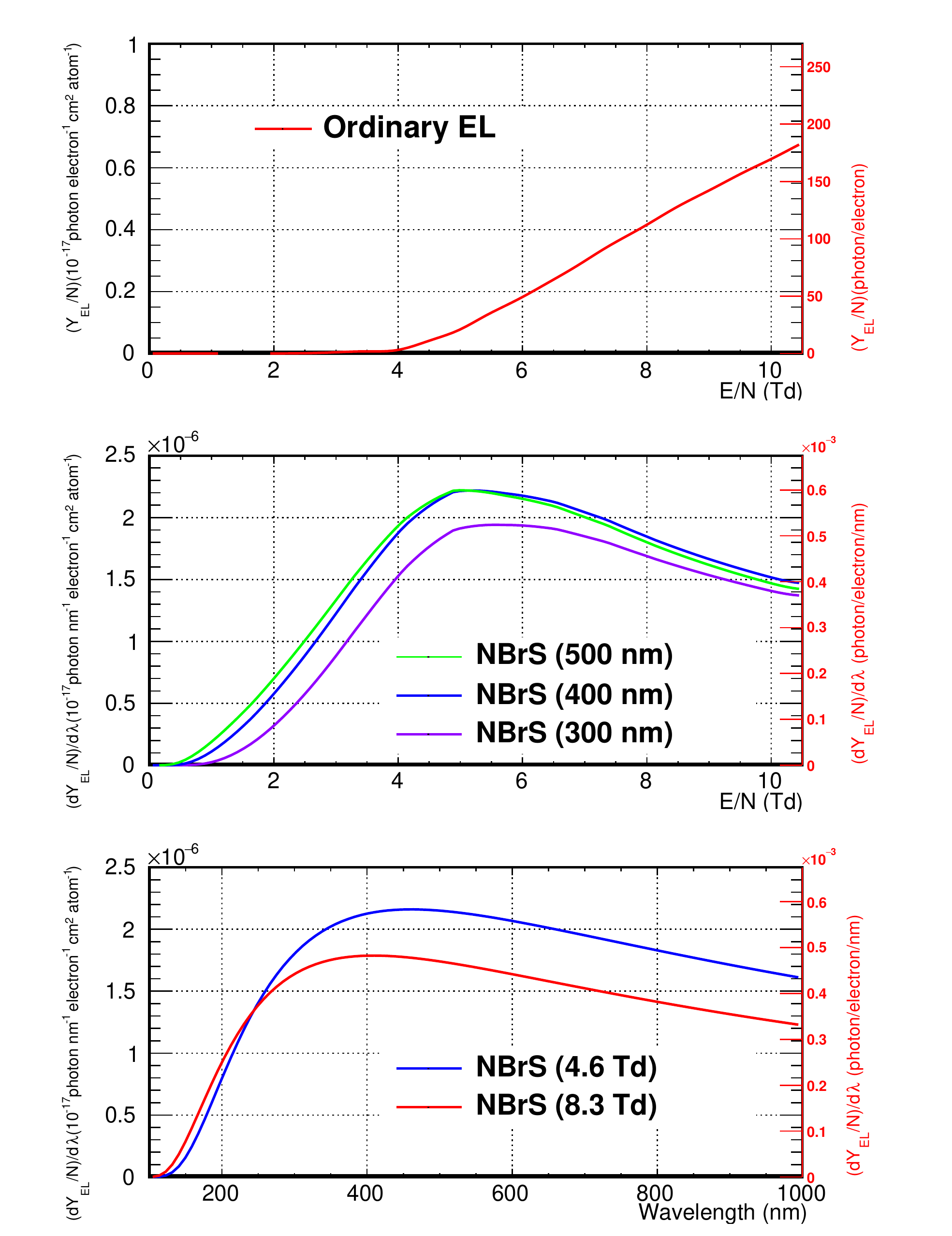}
  \caption{Theoretical emission light yield as a function of reduced electric field $E/N$ for the ordinary electroluminessence (EL) model (top) and the neutral bremsstrahlung (NBrS) model (middle). Emission light yield as a function of wavelength for the NBrS model (bottom). The axis labels on the right shown represented by red characters represent the light yield at room temperature (300 K), normal pressure (1 bar), and an electron drift distance of 1 cm.}
\label{Fig:theo}
\end{figure}

The EL process was explained by the ordinary mechanism, but without VL emissions. Recent measurements, however, demonstrated the presence of VL components in the EL of GAr, which was explained by Buzulutskov et al. \cite{Buzulutskov:2018vgg} using a new mechanism called neutral bremsstrahlung (NBrS). Because VL photons are easier to detect using photosensors than VUV photons, the use of VL components in particle detectors has been actively discussed \cite{Bondar:2019gjo,Aalseth:2020zdm}, and a detailed understanding of the wavelength spectrum is important for such applications. Figure~\ref{Fig:theo} illustrates the results of theoretical calculations of the EL emission yields as a function of a reduced electric field ($E/N$) at room temperature (300 K) for the ordinary EL model (top) \cite{Oliveira:2011xx} and the NBrS model (middle) \cite{Buzulutskov:2018vgg}. Although ordinary EL lights emit only above the 4-Td threshold, NBrS light also emits below it. The theoretical calculations of the NBrS light wavelength spectra are depicted in the bottom plot of Figure~\ref{Fig:theo} \cite{Buzulutskov:2018vgg}. The spectrum is a continuous distribution from 200 to 1,000 nm, which is different from that of ordinary EL light.

As shown in Fig.~\ref{Fig:mech}, VL emissions (300--450 nm) for the Ar-nitrogen gas mixture due to the nitrogen excimer are given by the following process \cite{Takahashi:1983}:
\begin{eqnarray}
  \mathrm{Ar^* + N_{2}}& \rightarrow &\mathrm{Ar + N_{2}^{*}(C^{3}\Pi_{u})},\\
  \mathrm{N_{2}^{*}(C^{3}\Pi_u)}&\rightarrow&\mathrm{N_{2}^{*}(B^{3}\Pi_g)+h\nu~(300-450~nm)}.
\end{eqnarray}
Nitrogen emission has a similar $E/N$ dependence as the ordinary EL with a 4-Td threshold, as it occurs due to the transition of Ar$^{*}$. It is known that more than 1 ppm of nitrogen impurity in liquid Ar degrades the Ar scintillation light \cite{WArP:2008rgv}. Thus, the nitrogen impurity inside the liquid phase is usually controlled at less than 1 ppm for liquid Ar double-phase detectors. However, the nitrogen impurity in the gas phase can be higher because the boiling point of N (77 K) is less than that of Ar (87 K), and nitrogen impurities tend to concentrate in the gas phase after a certain amount of time. Thus, the effect of nitrogen impurities should be carefully considered in order to understand the VL components.

In this study, detailed measurements of the wavelength spectrum of GAr EL in the VL region (300--600 nm) were conducted under electric fields up to $E/N$ = 10 Td at room temperature (300 K) and normal pressure (1 bar) using a spectrometer. Then, the spectra obtained using ordinary EL and NBrS and nitrogen impurity are interpreted. Note that these emission processes are described as functions of a reduced electric field $E/N$ when they are weakly associated with temperature and pressure at the same $E/N$. Thus, the results of the measurement at room temperature and normal pressure are also applicable to the liquid Ar temperature (87 K) for the same $E/N$. $E/N$ is defined as the electric field per molecular density of the Ar gas. The molecular density is inversely proportional to the temperature. $E/N$ at temperatures $T_1$ to $T_2$ is given by $(E/N)_{T_1}/(E/N)_{T_2}=T_2/T_1$ with the same electric field and pressure. Thus, 1 Td at 87 K is equivalent to 3.4 Td at 300 K with the same pressure and electric field.

\section{Experimental Apparatus}

\begin{figure}[!h]
  \centering\includegraphics[width=0.6\columnwidth]{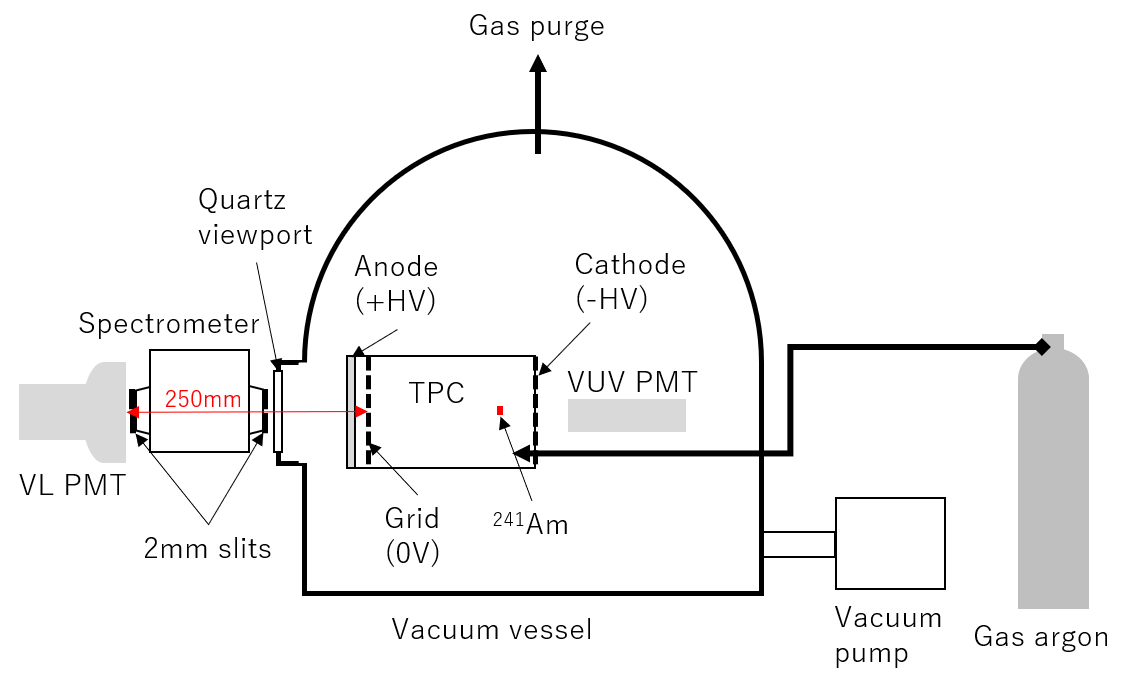}
  \centering\includegraphics[width=0.8\columnwidth]{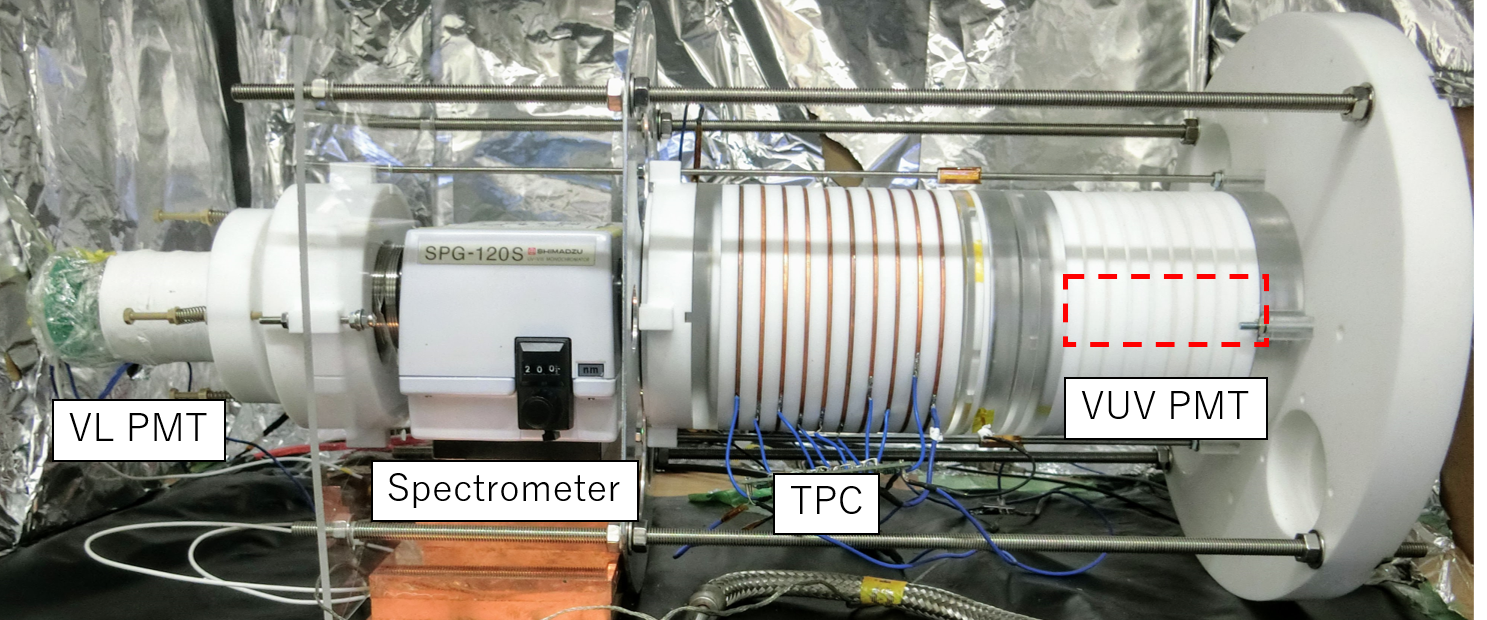}
  \caption{Schematic of the experimental setup (top). Picture of the TPC, spectrometer, and VL and VUV photomultiplier tubes (PMTs) (bottom).}
  \label{Fig:setup}
\end{figure}

Figure~\ref{Fig:setup} shows a schematic of the experimental setup of a GAr TPC with a fiducial volume of length 130 mm and diameter 64 mm. The TPC was constructed with polytetrafluoroethylene (PTFE) and Cu field shaping rings. No wavelength shifter was used to measure pure Ar luminescence. The grid was placed 10 mm away from the anode, after which the anode and cathode were used to generate luminescence and drift fields, respectively. The anode was transparent indium-tin-oxide (ITO)-coated quartz, which had a transmittance of more than 95\% for VL ($>$300 nm). An $\alpha$-ray source ($^{241}$Am, 300 Bq) was placed inside the TPC for signals.

\begin{figure}[!h]
\centering\includegraphics[width=0.9\columnwidth]{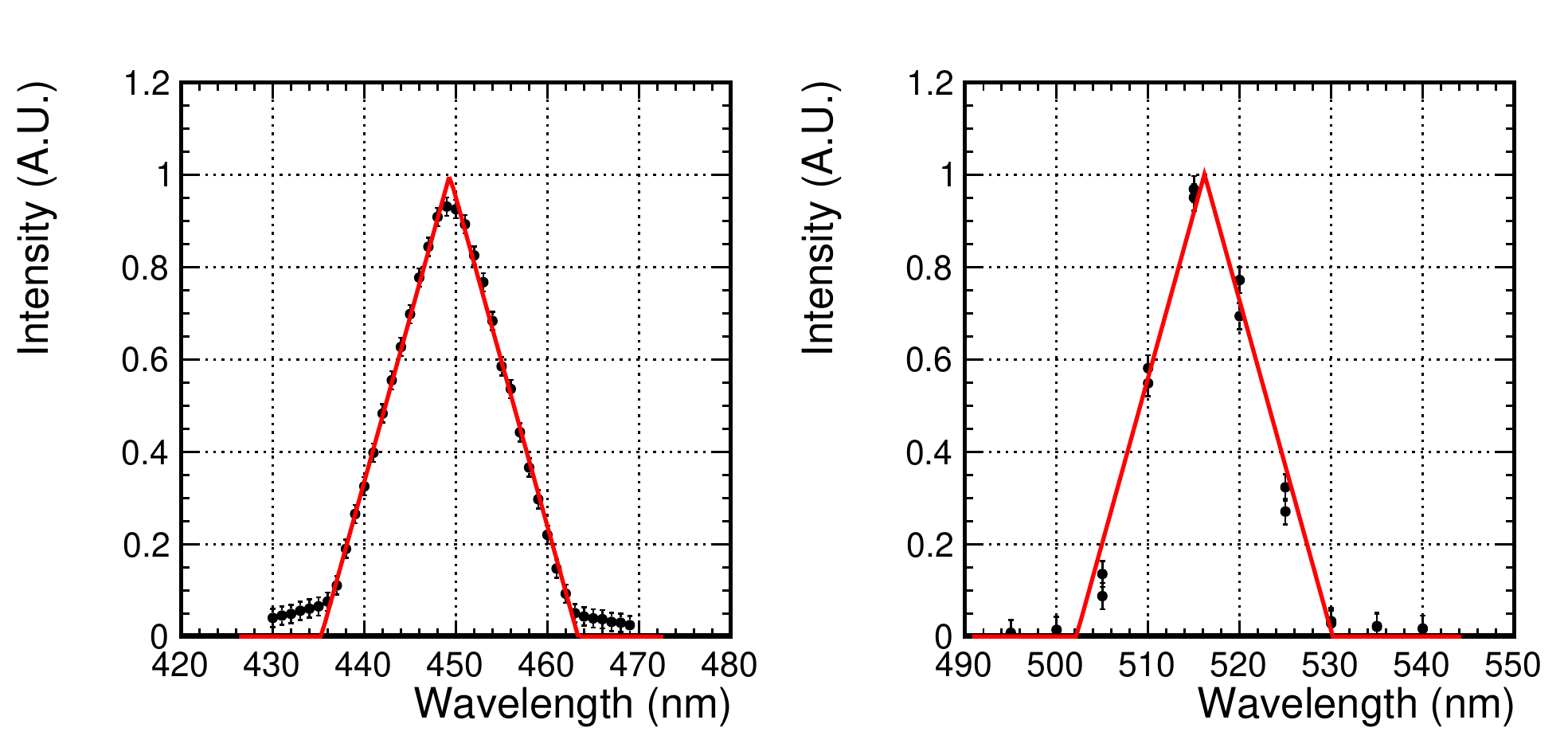}
\caption{Wavelength responses of the spectrometer for the 450 (left) and 520 nm lasers (right). Data are fitted to triangular functions (red lines).}
\label{Fig:laser}
\end{figure}

The TPC was placed inside a vacuum vessel with a quartz viewport. A spectrometer (Shimadzu Corporation SPG-120S) was placed outside the vessel near the viewport. We attached 2 mm slits to both the inlet and outlet of the spectrometer. The distance between the anode and the outlet slit was approximately 250 mm. Figure~\ref{Fig:laser} shows the responses of the spectrometer to monochromatic laser light with wavelengths of 450 nm (left) and 520 nm (right). These responses were approximately modeled by triangular functions $T(\lambda)$ around the peak wavelength $\lambda_{peak}$,
\begin{equation}
  T(\lambda) = \left\{
  \begin{array}{ll}
    A\left[1-\frac{|\lambda-\lambda_{peak}|}{w}\right]\;\;\; &(|\lambda-\lambda_{peak}|<w) \\
    0  &(|\lambda-\lambda_{peak}|>w)
  \end{array} \right.,
\label{Eq:sp}
\end{equation}
where the width of the function $w$ = 14 nm, and the height of the peak $A$ = 1. We observed a 5 nm shift in the responses caused by the combination of performances of the spectrometer and input laser. This shift was a source of systematic uncertainties observed in the measurements.

\begin{figure}[!h]
  \centering\includegraphics[width=0.6\columnwidth]{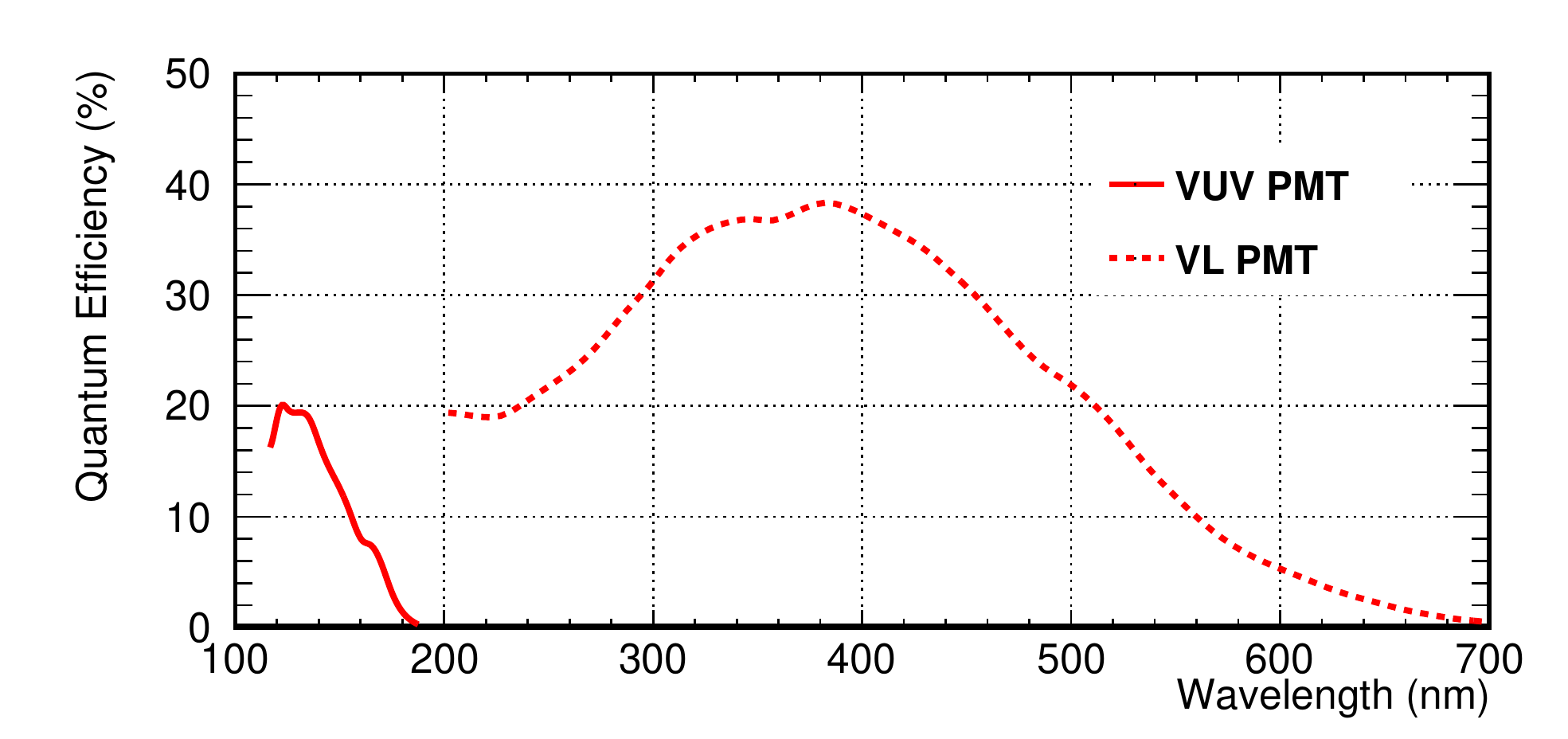}
  \caption{Quantum efficiencies of the two PMTs (VUV PMT and VL PMT) used in this measurement as a function of the wavelength.}
  \label{Fig:pmtqe}
\end{figure}

Two PMTs with different wavelength sensitivities were arranged near the spectrometer and cathode. The PMT near the spectrometer (Hamamatsu-R11065 ``VL PMT'' in this study), which had a quartz window, was sensitive to UV and VL, while that near the cathode (Hamamatsu-R6835 called ``VUV PMT''), which had a MgF$_2$ window, was only sensitive to VUV (Fig.~\ref{Fig:pmtqe}). A flash analog-to-digital converter (ADC; Struck Innovative Systeme SIS3316-250-14) with a sampling rate of 250 Ms/s and a resolution of 14 bits was used to read the PMT signals.

The vacuum vessel was depressurized to $10^{-2}$ Pa using a molecular turbo pump for at least one day before each measurement. Subsequently, the detector was filled with GAr at room temperature and normal pressure. The gas was pumped at a constant flow rate of 10 L/min during the experiment to avoid impurities from outgassing. To evaluate the effect of nitrogen impurities, three admixtures of Ar and nitrogen gases (G1, N10, and N100) were used in this measurement. The specifications of the gases are listed in Table~\ref{Tab:gas}.

\begin{table}[!h]
  \caption{Specification of Ar gases used in the measurements.}
  \centering\begin{tabular*}{0.6\columnwidth}{lll} \hline
  Gas type & N$_{2}$ composition & Other impurities\\ \hline
  G1 & $<$ 0.3 ppm &$<$ 0.1 ppm \\ \hline
  N10 & 10$\pm$1 ppm &$<$ 0.1 ppm \\ \hline
  N100 & 100$\pm$10 ppm &$<$ 0.1 ppm \\ \hline
  \end{tabular*}
  \label{Tab:gas}
\end{table}

Three datasets were used in this study.
\begin{itemize}
\item Data for wavelength spectrum\\
  High-purity G1 gas was used to obtain these data. The spectrometer wavelength was scanned from 240 to 660 nm with a 20 nm pitch at two luminescence fields of 4.6 and 8.3 Td. These $E/N$ values were chosen to match the measurements and theoretical calculations in \cite{Buzulutskov:2018vgg}.
\item Data for electric field dependence\\
  High-purity G1 gas was used to obtain these data. The luminescence field was scanned from 1 to 8 Td with 1-Td pitch at 300, 400, and 500 nm. Additional data for the 300 nm wavelength were obtained at 10 Td.
\item Data for nitrogen effect\\
  N10 and N100 gases were used for acquiring these data. The spectrometer wavelength was scanned from 240 to 660 nm with a 20-nm pitch at a luminescence field of 8.3 Td. The spectrometer wavelength for the N100 gas was scanned from 300 to 450 nm with a 2.5-nm pitch. Six additional wavelength datasets (316, 337, 358, 381, 406, and 434 nm) corresponding to the resonant wavelength of nitrogen emission were taken for the G1 and N10 gases \cite{Takahashi:1983}.
  
\end{itemize}

Note that 1 Td of the reduced electric field in the luminescence region (room temperature, normal pressure, and 1 cm luminescence region) corresponds to 245 V/cm.

\section{Results}

The primary scintillation light (S1) and ionization electrons were produced by the interaction between the $\alpha$-ray and the GAr atoms. The typical flight distance of the $\alpha$-ray was 3 cm. The ionization electrons drifted in the direction of the grid under a fixed drift field of 100 V/cm (0.4 Td). The electrons emitted secondary EL lights (S2) when they were in the 1 cm gap between the grid and the anode under a luminescence field (1--10 Td).

\begin{figure}[!h]
  \centering\includegraphics[width=0.8\columnwidth]{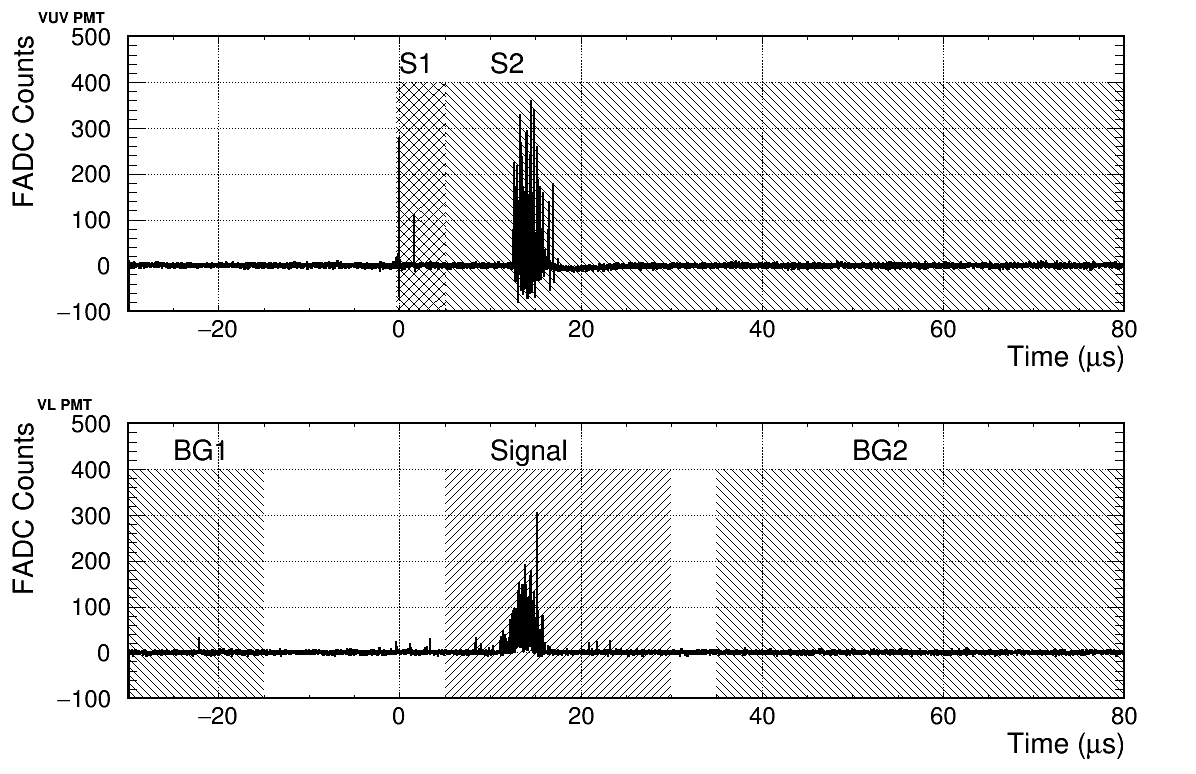}
  \caption{Typical waveform distributions of the VUV PMT (top) and VL PMT (bottom) for the setup without a spectrometer.}
  \label{Fig:waveform}
\end{figure}

The typical waveform distributions measured by the Flash ADC are illustrated in Fig.~\ref{Fig:waveform}. The top and bottom plots correspond to the VUV PMT and VL PMT signals, respectively. These distributions were obtained without a spectrometer. The actual light yield of the VL PMT with a spectrometer was less than one photoelectron (PE). The sharp peak for the VUV PMT at t = 0 corresponds to the S1 signal, while the broad peak at approximately t = 15 $\mu$s corresponds to the S2 signal.

\begin{table}[!h]
  \caption{Definition of light yield variables.}
  \centering\begin{tabular*}{0.6\columnwidth}{lll} \hline
  Name & PMT& Interval of integration\\ \hline
  S1 & VUV PMT &[$-$0.4 $\mu$s, 5 $\mu$s]\\
  S2 & VUV PMT &[5 $\mu$s, 80 $\mu$s]\\
  Signal & VL PMT &[5 $\mu$s, 30 $\mu$s]\\
  BG1 & VL PMT &[$-$30 $\mu$s, $-$15 $\mu$s]\\
  BG2 & VL PMT &[35 $\mu$s, 80 $\mu$s]\\
  \hline
  \end{tabular*}
  \label{Tab:para}
\end{table}

\begin{figure}[!h]
  \centering\includegraphics[width=0.6\columnwidth]{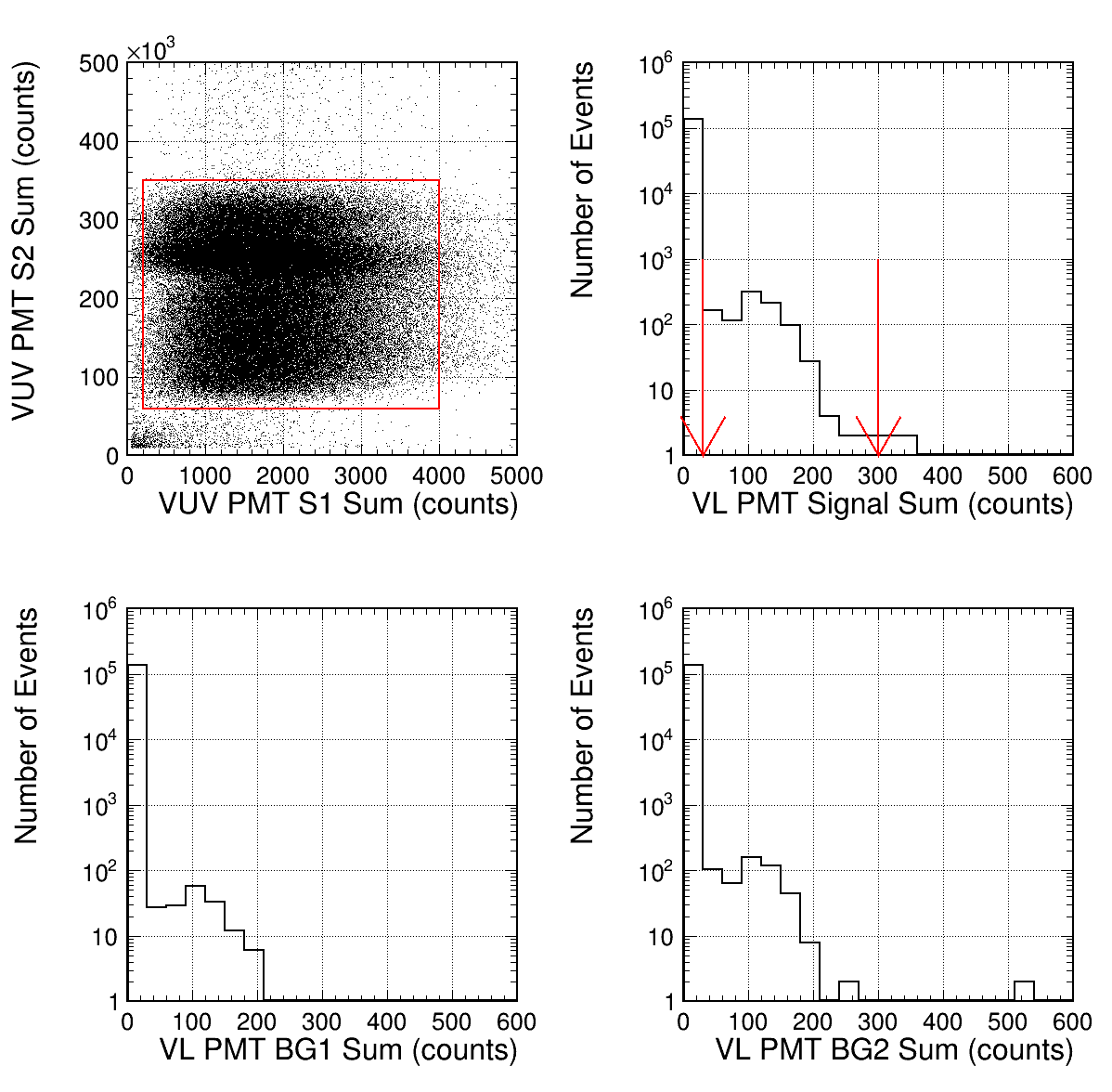}
  \caption{Integrated light yield distributions of the VUV PMT waveform in the S1 and S2 regions (top-left). Data for the plot are obtained at a spectrometer wavelength of 400 nm and a luminescence field of 8.3 Td using the G1 gas. Events in the red box are selected for the wavelength spectrum calculation. The integrated light yield distribution of the VL PMT waveform is in the Signal (top right), BG1 (bottom left), and BG2 (bottom right) regions.}
  \label{Fig:s1s2}
\end{figure}

Five light yields (S1, S2, Signal, BG1, and BG2) were calculated by integrating the waveform distributions of the VUV and VL PMT, as illustrated in Table~\ref{Tab:para}. The integration intervals are indicated by the hatched regions in Fig.~\ref{Fig:waveform}. Events were selected using S1 and S2 in the red box in the top-left plot in Fig.~\ref{Fig:s1s2}. Data for the plot were obtained at a spectrometer wavelength of 400 nm and a luminescence field of 8.3 Td using G1 gas, and 137,684 events were selected (N$_0$). Distributions of Signal, BG1, and BG2 for the selected events are depicted in the top-right, bottom-left, and bottom-right plots of Fig.~\ref{Fig:s1s2}, respectively. As mentioned earlier, no photons were observed most of the time, and the peaks around 100 counts corresponded to single PE events. The VL PMT was operated with gain for single PE events of approximately 130 counts, and the PMT had distinct separation between the noise level and the single PE events \cite{Kimura:2020mpd}. The single PE events were selected based on the light yields within the range of 30--300 counts, and the selected events were N$_\mathrm{Signal}$ = 943, N$_\mathrm{BG1}$ = 166, and N$_\mathrm{BG2}$=502 for Signal, BG1, and BG2, respectively.
The signal region contained background events, mainly due to the accidental coincidence of the PMT dark counts. Background contamination was estimated using BG1 and BG2 normalized to the window width,  $\mathrm{N_{Background}}$ = $(25~\mu s /60~\mu s)\times\mathrm{(N_{BG1}+N_{BG2})}$ = 278, and the number of events after background subtraction was $\mathrm{N_{Signal}}$ $-$ $\mathrm{N_{Background}}$ = 665. Assuming that the light yield followed a Poisson distribution, the light yield in the unit of PEs equaled the fraction of events in the single PE peak is $\mathrm{(N_{Signal}}$ $-$ $\mathrm{N_{Background})/N_{0}}$ = 0.00483. 

\begin{figure}[!h]
  \centering\includegraphics[width=0.8\columnwidth]{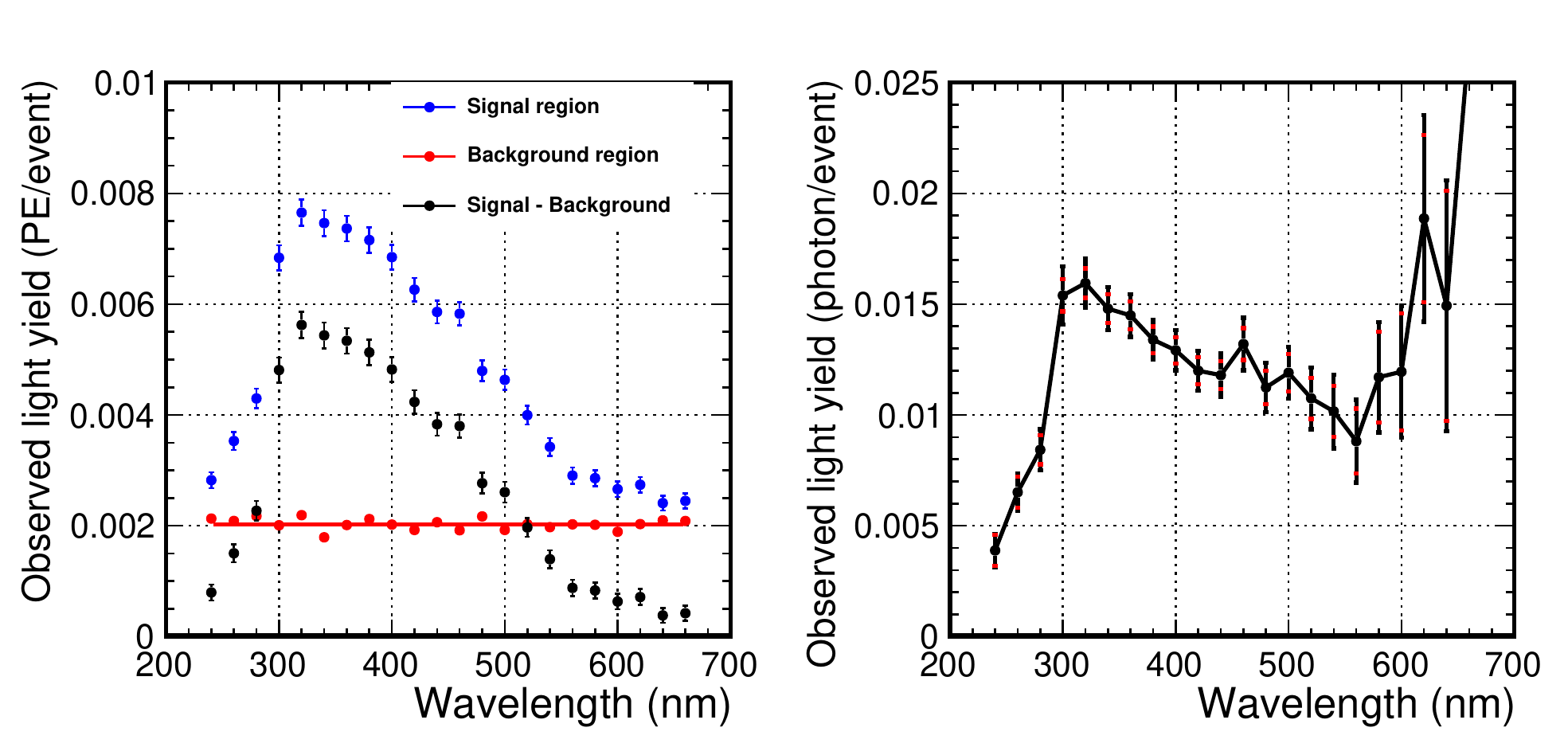}
  \caption{Wavelength spectra for G1 gas at a luminescence field of 8.3 Td. Left: Observed light yields in the unit of photoelectrons as a function of the wavelength in the Signal region (blue points), BG1 + BG2 region (red points), and background-subtracted signal (black points). Right: Wavelength spectrum after correcting for PMT quantum efficiency.}
  \label{Fig:sb}
\end{figure}

By repeating the same analysis steps for different configurations (spectrometer wavelength, luminescence field, and gas type), we obtained the desired spectrum. The left plot in Fig.~\ref{Fig:sb} illustrates the obtained wavelength spectra using G1 gas at a luminescence field of 8.3 Td. The blue, red, and black points denote the light yields in the signal, background, and signal--background region, respectively. As the light yield in the background region was steadily constant, the average value (red line) was used for subtraction. Finally, the wavelength spectrum (right plot in Fig.~\ref{Fig:sb}) was obtained after correcting for the VL PMT quantum efficiency in Fig.~\ref{Fig:pmtqe}).
Two sources of systematic uncertainties were considered for the observed light yield.
\begin{itemize}
\item PMT quantum efficiency.\\
  As discussed in the previous section, the wavelength of light after passing through the spectrometer followed the distribution described by Eq.~\ref{Eq:sp}, and a systematic shift of $\pm$5 nm was observed for the spectrometer. Thus, the wavelength to be used to calculate the quantum efficiency of the VL PMT is uncertain. The difference in quantum efficiency when changing the wavelength $\pm$7 nm was assigned as the systematic uncertainty of the observed light yield.
\item Reproducibility of measurement\\
  The stability of measurement during the data-acquisition period was estimated by repeating the measurement. A relative uncertainty of 5\% was assigned to the observed light yield.
\end{itemize}
The red inner error bars in the right plot of Fig~\ref{Fig:sb} demonstrate statistical uncertainty, and the outer error bars with black color represent the total ($\sqrt{\mathrm{statistical^2+systematic^2}}$) uncertainty.

\begin{figure}[!h]
\centering\includegraphics[width=0.8\columnwidth]{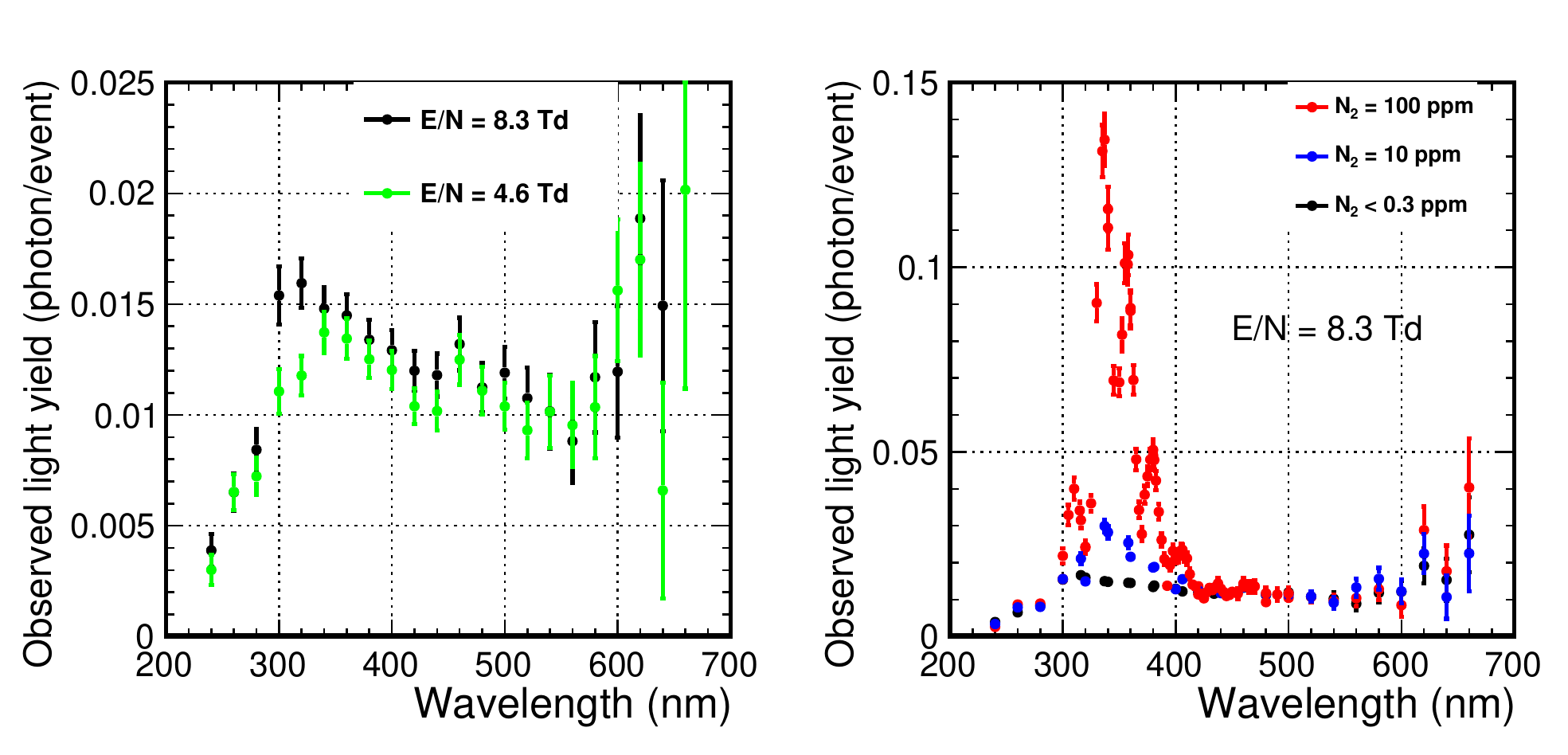}
\caption{Wavelength spectra after correcting the PMT quantum efficiency. Left: Spectra for high-purity GAr with electric fields of 4.6 Td (green points) and 8.3 Td (black points). Right: Spectra for GAr with N concentrations of less than 0.3 ppm (black points), 10 ppm (blue points), and 100 ppm (red points).}
\label{Fig:rawc}
\end{figure}

The wavelength spectra obtained are shown in Fig.~\ref{Fig:rawc}. The left plot shows the spectra for the G1 gas with electric fields of 4.6 Td (green points) and 8.3 Td (black points). The right plot shows the spectra for the G1 gas (black points), N10 gas (blue points), and N100 gas (red points) at $E/N$ = 8.3 Td.
The transmittance of the ITO-quartz degraded for the wavelength below 300 nm, and there were background contributions from the secondary light of the spectrometer for the wavelength above 600 nm (300 nm$\times$2). Thus, a wavelength range of 300--600 nm was used for the rest of the analysis.

\section{Discussion}
\subsection{Electric Field Dependence}

\begin{figure}[!h]
\centering\includegraphics[width=0.8\columnwidth]{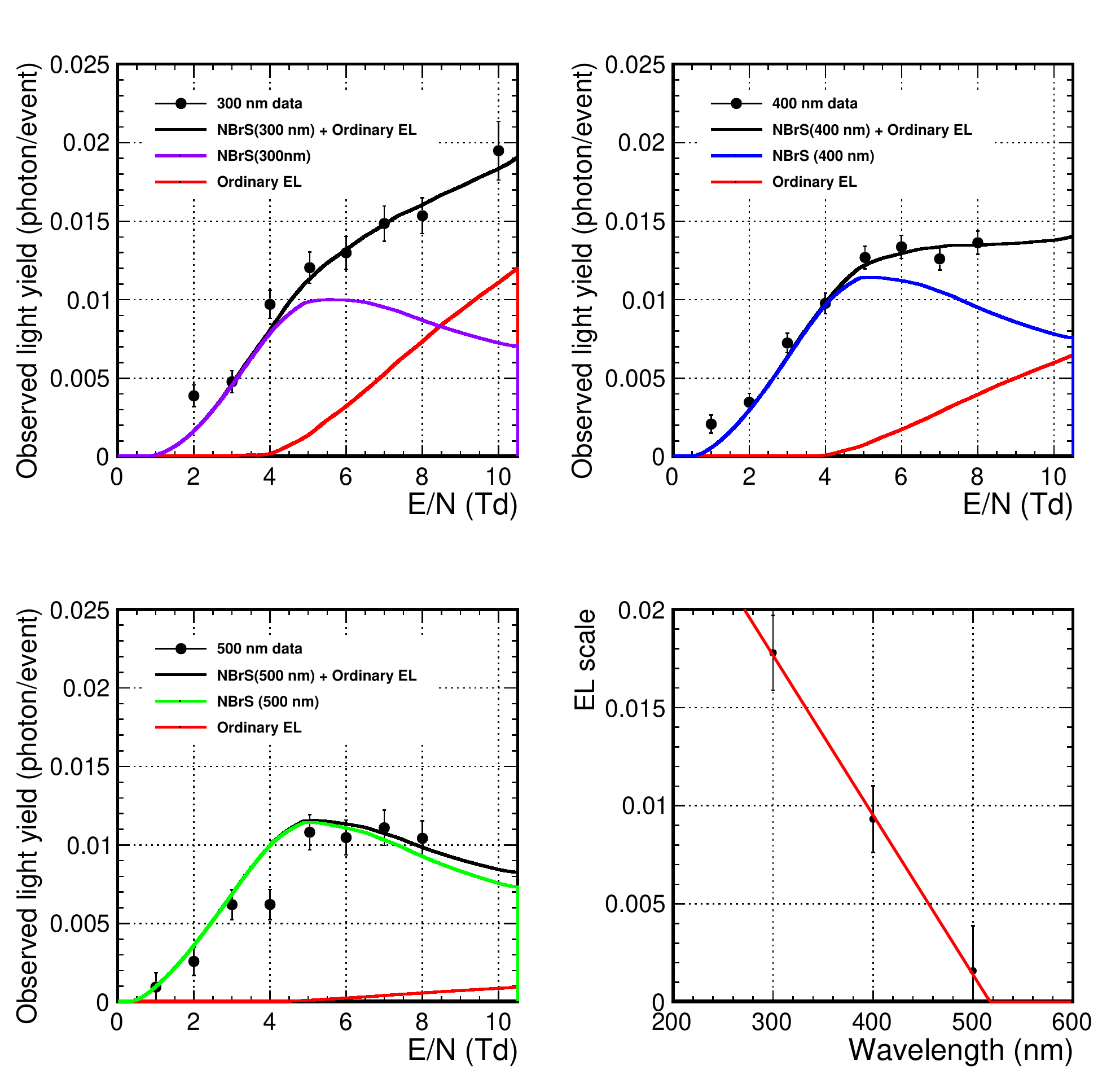}
\caption{Observed light yield as a function of $E/N$ at wavelengths of 300 (top left), 400 (top right), and 500 nm (bottom left). The three distributions are simultaneously fitted to the sum of the model functions (black) of NBrS (purple, blue, and green) and ordinary EL (red). The bottom-right plot shows the scale factors for the ordinary EL as a function of wavelength.}
\label{Fig:v}
\end{figure}

Figure~\ref{Fig:v} illustrates the results of the electric field dependence data at three wavelengths. The wavelength around each nominal wavelength had an interval of approximately 14 nm (width $w$ of the spectrometer in Eq~(\ref{Eq:sp})). Light rays were observed in low electric fields below 4 Td. The light yields appeared saturated at higher electric fields. Thus, the ordinary EL model is not enough to explain these distributions. To quantitatively estimate the contribution of the NBrS model, the data points were simultaneously fitted to the following function,
\begin{equation}
  F_\mathrm{Ar}(\lambda,E/N)  = S_\mathrm{EL}(\lambda)F_\mathrm{EL}(E/N)+ S_\mathrm{NBrS}F_\mathrm{NBrS}(\lambda,E/N).
  \label{Eq:g1}
\end{equation}
$F_\mathrm{EL}(E/N)$ and $F_\mathrm{NBrS}(\lambda,E/N)$ are the light yield functions for the ordinary EL and NBrS models in Fig.~\ref{Fig:theo}, respectively. The scale factors, $S_\mathrm{EL}(\lambda)$ and $S_\mathrm{NBrS}$, were determined by the fit.

\begin{table}[!h]
  \caption{Scale factors obtained by fitting the data points to the electric field dependence of the observed light yield distribution.}
  \centering\begin{tabular*}{0.7\columnwidth}{ll} \hline
  Fit parameter & Value\\ \hline
  $\chi^2$/ndf& 41.9/25\\ \hline
  $S_\mathrm{NBrS}$&$(5.1\pm0.2)\times10^{3}$\\ 
  $S_\mathrm{EL}$(300 nm)& $(1.81\pm0.22)\times10^{-2}$\\
  $S_\mathrm{EL}$(400 nm)& $(0.97\pm0.22)\times10^{-2}$\\
  $S_\mathrm{EL}$(500 nm)& $(0.19\pm0.25)\times10^{-2}$\\\hline
  $S_{0}$& $-(8.1\pm1.5)\times10^{-5}$\\
  $\lambda_{0}$& $(521\pm31)$\\
  \hline
  \end{tabular*}
  \label{Tab:vfit}
\end{table}

The fit results are summarized in Table~\ref{Tab:vfit}, and the fitted functions in Figure~\ref{Fig:v}. The $\chi^2$/ndf of the fit was 41.9/25, which is marginally worse than that of the fitted function because two data points (300 nm with 2 Td  and 500 nm with 4 Td) were significantly different from those of the fitted function. Note that the result did not change significantly, even after excluding the two points from the fit. The bottom-right plot in Figure~\ref{Fig:v} shows $S_\mathrm{EL}$ as a function of wavelength. The scale factor at 300 nm was the largest because of the significant UV contribution from the third continuum emission in Eq.~(\ref{Eq:uv}). However, the scale factor at 500 nm was consistent with zero within its uncertainty. Although the spectrum of the ordinary EL emission is not entirely understood, the wavelength dependence of the scale factors is approximately a straight line,
\begin{equation}
  S_{EL}(\lambda) =\left\{
  \begin{array}{ll}
    S_{0}(\lambda-\lambda_{0}) \;\;\;\;\; &(\lambda < \lambda_0)\\
    0 \;\;\;\;\; &(\lambda > \lambda_0)\\
  \end{array}  \right..
  \label{Eq:sl}
\end{equation}
The parameters $S_{0}$ and $\lambda_0$ were obtained by fitting the data, as shown in Table~\ref{Tab:vfit}.

\subsection{Emission Light Yield of the NBrS Model}
In Figure~\ref{Fig:theo}, the theoretical prediction of the emission light yield ($F_\mathrm{NBrS}$) is equal to  $2.2\times10^{-6} (10^{-17}\mathrm{photon/electron/cm^2/atom})$ for an $E/N$ of 4.6 Td and a wavelength of 500 nm. The predicted number of photons emitted by the NBrS mechanism ($N_\gamma^{emit}$) for this setup, obtained using $^{241}$Am $\alpha$-ray and a spectrometer, was calculated using $F_\mathrm{NBrS}$ as follows:
\begin{equation}
  N_\gamma^{emit} = N_e \times \rho \times d \times w \times F_\mathrm{NBrS} = 1.7\times10^{3}.
\end{equation}
where $N_e=E_\alpha/W_i$ = 5.49 MeV/26.4 eV $=$ 2.1$\times10^{5}$ \cite{Aprile:2008bga} is the number of drift electrons, $\rho=2.7\times10^{19}$ atom/cm$^{3}$ is the number density of the Ar atom, $d=1$ cm is the distance of the luminescence field, and $w=$14 nm is the width of the spectrometer response in Eq.~(\ref{Eq:sp}). In contrast, in Figure~\ref{Fig:rawc}, the number of observed photons $N_{\gamma}^{obs}=1.1\times10^{-2}$ for $E/N$ = 4.6 Td and wavelength = 500 nm. Thus, the overall photon detection efficiency was calculated as follows:
\begin{equation}
  \epsilon=N_{\gamma}^{obs}/N_\gamma^{emit}= 1.1\times10^{-2}/1.6\times10^{3}=6.5\times10^{-6}.
\end{equation}
  The efficiency was mainly ascribed to the geometrical acceptance ($A_\mathrm{geo}$), which can be calculated using the distance from the luminescence region to the outlet slit (25 cm), and the size of the slit (2 mm$\times$5 mm).
\begin{equation}
  A_\mathrm{geo} = \frac{\mathrm{2~mm} \times \mathrm{5~mm}}{4 \pi \times (\mathrm{25~cm})^2} = 1.3 \times 10^{-5}.
\end{equation}
$A_\mathrm{geo}$ is two times larger than $\epsilon$, which can be approximately explained by the transmittance of the ITO quartz and the quartz viewport (approximately 95\% each), and the efficiency of the spectrometer ($>$50\%~\cite{SPG120}). Thus, the absolute value of the observed light yield of this measurement was approximately consistent with the prediction of the NBrS model.

\subsection{Wavelength Spectrum}

\begin{figure}[!h]
\centering\includegraphics[width=0.6\columnwidth]{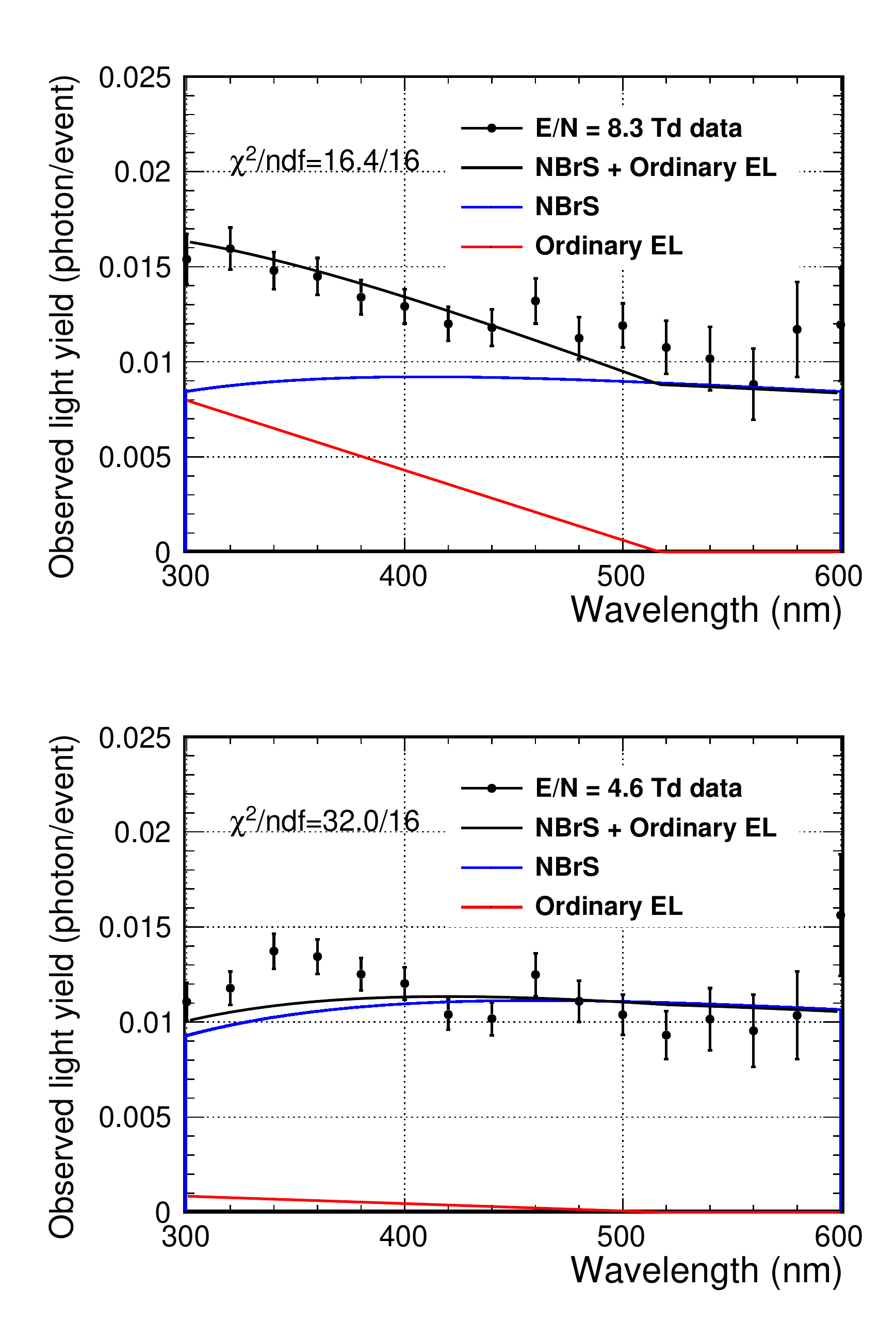}
\caption{Measured wavelength spectrum for $E/N$ = 8.3 (top) and 4.6 Td (bottom). The spectra for NBrS and ordinary EL are overlaid.}
\label{Fig:w}
\end{figure}

Because the scale factors for the ordinary EL model ($S_\mathrm{EL}$) and the NBrS model ($S_\mathrm{NBrS}$) help predict all wavelengths mentioned in the previous section, the model function of Eq.~(\ref{Eq:g1}) explains the wavelength spectrum. Figure~\ref{Fig:w} depicts an overlay of the model functions to the wavelength spectrum data at the reduced electric fields of 8.3 Td (top) and 4.6 Td (bottom).
  The $\chi^2/ndf$ values between the data points and the model functions were 16.4/16 and 32.0/16 for the 8.3 and 4.6 Td data, respectively. The spectrum data and model functions were in good agreement, except for two data points (340 and 360 nm with 4.6 Td). In this analysis, the spectrum for ordinary EL emission was modeled using a straight line (Eq.~(\ref{Eq:sl})). However, the spectrum of G1 gas with 8.3 Td (Fig.~\ref{Fig:rawc}) showed a significant discontinuity below 300 nm, which points to the existence of a resonant-type structure for the ordinary EL emission. A finer-wavelength scan with more statistics will be required for a detailed explanation of ordinary EL emission. The spectrum at 4.6 Td was explained by the nearly pure NBrS emission, and the contribution of the EL emission was less than 10\%.

\subsection{Nitrogen Effect}

\begin{figure}[!h]
  \centering\includegraphics[width=0.8\columnwidth]{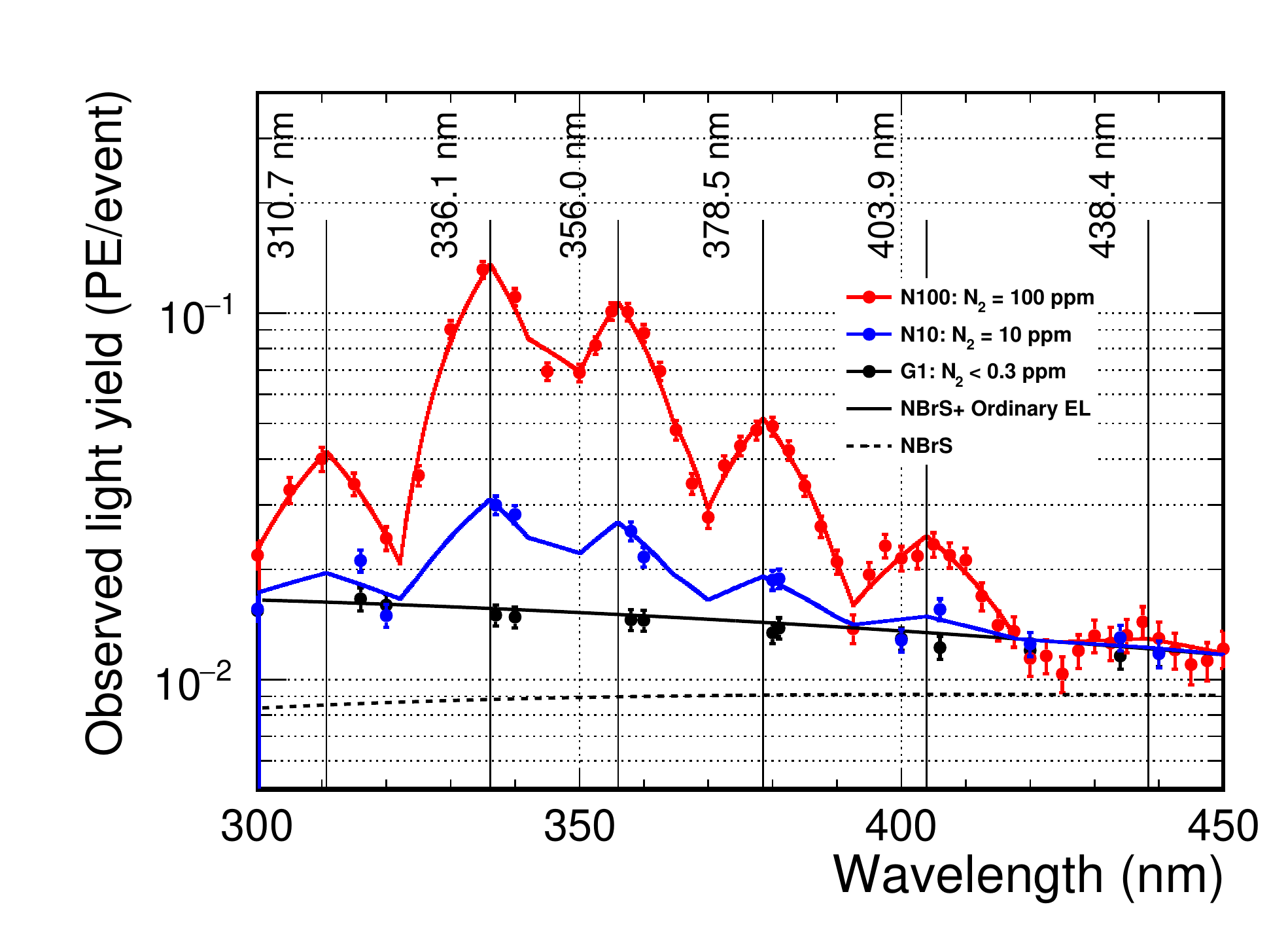}
  \caption{Wavelength spectra for N$_2$ concentrations of $<$0.3 ppm (black points), and 10 ppm (blue points), and 100 ppm (red points).}
\label{Fig:n2}
\end{figure}

As mentioned earlier, a possible explanation for the VL component of the EL emission other than the NBrS model is the nitrogen excimer emission. This effect was estimated using an Ar--N gas mixture. The right hand-side plot in Figure~\ref{Fig:rawc} shows a significant excess between 300 and 400 nm in the N100 data due to nitrogen. However, the N100, N10, and G1 wavelength spectra are consistent above a wavelength of 450 nm. Thus, we assumed that there were no significant nitrogen contributions above 450 nm. Figure~\ref{Fig:n2} shows the wavelength spectra from 300 to 450 nm with an electric field of 8.3 Td using G1 gas (black points), N10 gas (blue points), and N100 gas (red points), respectively. The model curves  of NBrS (dotted black line) and NBrS + ordinary EL (solid black line) are shown (the same curves as in Fig.~\ref{Fig:w}).
The N100 gas data had six resonant structures consistent with the measurements reported by Takahashi et al.\cite{Takahashi:1983}. The peak wavelengths ($\lambda_{peak,i}$) and heights ($A_i$) of the resonances were obtained by fitting the N100 spectra using a model function:
\begin{eqnarray}
F_\mathrm{N100}(\lambda)=F_\mathrm{Ar}(\lambda)+\sum_{i=1}^{6}T_{i}(\lambda),
\label{Eq:N100}
\end{eqnarray}
where $F_\mathrm{Ar}$ is the model function of Eq.~(\ref{Eq:g1}) (solid black line in Fig~\ref{Fig:n2}), and the triangular function $T_{i}$ is
\begin{equation}
  T_i(\lambda) = \left\{
  \begin{array}{ll}
    A_i\left[1-\frac{|\lambda-\lambda_{peak,i}|}{w}\right]\;\;\; &(|\lambda-\lambda_{peak,i}|<w) \\
    0  &(|\lambda-\lambda_{peak,i}|>w)
  \end{array} \right.,
\end{equation}
where the width of the function $w$ is fixed at 14 nm.
The fit results are summarized in Table~\ref{Tab:n2}. As mentioned earlier, a systematic uncertainty of 5 nm was assigned to the wavelength considering the shift observed in Fig.~\ref{Fig:laser}.

\begin{table}[!h]
  \caption{Summary of the fit results of six observed emission peaks in the N100 data. The predicted wavelengths for the $N_{2}^{*}(C^{3}\Pi_g)\rightarrow N_{2}^{*}(B^{3}\Pi_g)+h\nu$ transition \cite{Takahashi:1983}.}
  \centering\begin{tabular*}{0.9\columnwidth}{lllll} \hline
  &\multicolumn{2}{c}{This work}&\multicolumn{2}{c}{Predicted\cite{Takahashi:1983}}\\
  $i$ &$\lambda_{peak,i}\pm_{fit}\pm_{sys}$& $A_i\pm_{fit}$  &V'=0 &V'=1\\
&   (nm)& $10^{-2}$(A.U.) &(nm) &(nm)\\
  \hline
1&  $310.7\pm0.6\pm5.0$ & $(2.6\pm0.2)$ &N/A&315.9\\ 
2&  $336.1\pm0.2\pm5.0$ & $(12.0\pm0.4)$ &337.1&333.9\\ 
3&  $356.0\pm0.1\pm5.0$ & $(9.3\pm0.2)$ &357.7&353.7\\ 
4&  $378.5\pm0.1\pm5.0$ & $(3.7\pm0.4)$ &380.5&375.5\\ 
5&  $403.9\pm0.5\pm5.0$ & $(1.1\pm0.2)$ &405.9&399.8\\ 
6&  $438.4\pm7.1\pm5.0$ & $(0.08\pm0.07)$ &434.4&427.0\\ 
  \hline
  \end{tabular*}
  \label{Tab:n2}
\end{table}

The wavelength spectrum with N10 gas was modeled using the following function,
\begin{eqnarray}
F_\mathrm{N10}(\lambda)=F_\mathrm{Ar}(\lambda)+\alpha\sum_{i=1}^{6}T_{i}(\lambda),
\end{eqnarray}
where $F_\mathrm{Ar}$ and $T_{i}(\lambda)$ have the same functions as those in Eqs.~(\ref{Eq:N100}). The scale, $\alpha$, was determined by fitting the data points to the wavelength spectrum Thus, $\alpha=0.12\pm0.01$ was obtained, which is consistent with the ratio of the nitrogen content of N10 ($10 \pm 1$ ppm) to N100 ($100 \pm 10$ ppm), i.e., 0.1. In addition, the G1 data were fitted to the model function,
\begin{eqnarray}
F_\mathrm{G1}(\lambda)=\beta F_\mathrm{Ar}(\lambda)+\alpha\sum_{i=1}^{6}T_{i}(\lambda),
\end{eqnarray}
and the fit result showed that $\alpha=0.001\pm0.007$ and $\beta=0.96\pm0.03$. Therefore, the residual nitrogen impurity in the G1 data was $0.1\pm0.7$ ppm, which is consistent with the uncertainty of 0 ppm.
In summary, a small amount of nitrogen impurities ($>$10 ppm) can cause VL emission from 300 to 450 nm. However, the G1 data of this measurement contained nitrogen impurities of less than 1 ppm, as well as negligible nitrogen emissions. As shown in Fig.~\ref{Fig:mech}, the nitrogen emission was produced from the transition of Ar$^{*}$, which requires at least 4 Td of the reduced electric field. Thus, nitrogen emissions cannot explain the VL emission below 4 Td. Even if small packets of VL emissions were produced from unknown impurities inside the Ar gas, it is highly unlikely that the emission occurred below 4 Td. Thus, NBrS emissions are the most reasonable model to explain the observed VL emission.
In addition, it is possible to utilize nitrogen emissions in the gas phase for the double-phase detector if the amount of nitrogen impurities in the liquid phase can be maintained below 1 ppm.

\section{Summary}

The GAr EL in the VL region (300 to 600 nm) was studied using the GAr TPC, with $^{241}$Am $\alpha$-rays as the signal source at room temperature and normal pressure. The secondary emission light waves from the TPC luminescence region were dispersed using a spectrometer. The wavelength spectrum and luminescence-field dependence of the light yield were compared with those of the ordinary EL and NBrS models. The effect of nitrogen impurities on the light yield was evaluated using the Ar-nitrogen mixture gas. We conclude that the ordinary EL model and nitrogen emission alone cannot explain the wavelength spectrum and electric field dependences of the observed light in the VL region. The inclusion of the NBrS model can enable a comprehensive explanation of the phenomena.

\section*{Acknowledgements}

The authors thank Dr. A.~Buzulutskov for helpful discussions on the NBrS model.

This work is part of the research performed under the Waseda University Research Institute for Science and Engineering (project number 2016A-507/2020N-006), supported by JSPS Grant-in-Aid for Scientific Research on Innovative Areas (15H01038/17H05204), Grant-in-Aid for Scientific Research(B) (18H01234), and Grant-in-Aid for JSPS Research Fellow (20J20839). The authors acknowledge the support of the Institute for Advanced Theoretical and Experimental Physics, Waseda University.

%% The Appendices part is started with the command \appendix;
%% appendix sections are then done as normal sections
%% \appendix

%% \section{}
%% \label{}

%% If you have bibdatabase file and want bibtex to generate the
%% bibitems, please use
%%
%%  \bibliographystyle{elsarticle-num} 
%%  \bibliography{<your bibdatabase>}

%% else use the following coding to input the bibitems directly in the
%% TeX file.

\end{document}